\documentclass[prl, twocolumn, superscriptaddress]{revtex4}
\usepackage{amssymb}
\usepackage{amsmath}
\usepackage{graphicx}
\usepackage{epsfig}
\usepackage{longtable}
\usepackage{epstopdf}
\usepackage{dcolumn}
\usepackage{bm}

\begin{document}

\title{Optical spin injection and spin lifetime in Ge heterostructures}
\author{F. Pezzoli}
\altaffiliation{fabio.pezzoli@unimib.it}
\affiliation{LNESS-Dipartimento di Scienza dei Materiali, Universit\`{a} degli Studi di Milano-Bicocca, I-20125 Milano, Italy}

%\affiliation{LNESS-Dipartimento di Scienza dei Materiali, Universit\`{a} degli Studi di Milano-Bicocca, Via R. Cozzi, I-20125 Milano, Italy}
%\affiliation{LNESS-Dipartimento di Fisica, Politecnico di Milano, Piazza Leonardo da Vinci 32, I-20133 Milano, Italy}

\author{F. Bottegoni}
\affiliation{LNESS-Dipartimento di Fisica, Politecnico di Milano, I-20133 Milano, Italy}

\author{D. Trivedi}
\affiliation{Department of Physics and Astronomy, University of Rochester, Rochester, NY 14627}

\author{F. Ciccacci}
\affiliation{LNESS-Dipartimento di Fisica, Politecnico di Milano, I-20133 Milano, Italy}

\author{A. Giorgioni}
\affiliation{LNESS-Dipartimento di Scienza dei Materiali, Universit\`{a} degli Studi di Milano-Bicocca, I-20125 Milano, Italy}

\author{P. Li}
\affiliation{Department of Electrical and Computer Engineering, University of Rochester, Rochester, NY 14627}

\author{S. Cecchi}
\affiliation{LNESS-Dipartimento di Fisica, Politecnico di Milano, I-20133 Milano, Italy}

\author{E. Grilli}
\affiliation{LNESS-Dipartimento di Scienza dei Materiali, Universit\`{a} degli Studi di Milano-Bicocca, I-20125 Milano, Italy}

\author{Y. Song}
\affiliation{Department of Physics and Astronomy, University of Rochester, Rochester, NY 14627}

\author{M. Guzzi}
\affiliation{LNESS-Dipartimento di Scienza dei Materiali, Universit\`{a} degli Studi di Milano-Bicocca, I-20125 Milano, Italy}

\author{H. Dery}
\affiliation{Department of Physics and Astronomy, University of Rochester, Rochester, NY 14627}
\affiliation{Department of Electrical and Computer Engineering, University of Rochester, Rochester, NY 14627}

\author{G. Isella}
\affiliation{LNESS-Dipartimento di Fisica, Politecnico di Milano, I-20133 Milano, Italy}

\begin{abstract}
We demonstrate optical orientation in Ge/SiGe quantum wells and study their spin properties. The ultrafast electron transfer from the center of the Brillouin zone to its edge allows us to achieve high spin-polarization efficiencies and to resolve the spin dynamics of holes and electrons. The circular polarization degree of the direct-gap photoluminescence exceeds the theoretical bulk limit, yielding $\sim$37\% and $\sim$85\% for transitions with heavy and light holes states, respectively. The spin lifetime of holes at the top of the valence band is found to be $\sim$0.5~ps and it is governed by transitions between heavy and light hole states. Electrons at the bottom of the conduction band, on the other hand, have a spin lifetime that exceeds 5~ns below 150~K. Theoretical analysis of the electrons spin relaxation indicates that phonon-induced intervalley scattering dictates the spin lifetime.
\end{abstract}


\maketitle

The spin degree of freedom in semiconductors is an attractive and promising means to perform both classical and quantum information processing in a scalable solid state framework \cite{Kane_Nature98,Wolf_science01,Zutic_RMP04,Fodor_JPCM06,Dery_Nature07}. Compared to other materials systems, group IV compounds have a greater potential in this field both because of the long spin lifetime \cite{Appelbaum_Nature07,Liu_NanoLett10,Han_PRL10,Guite_PRL11}, and because they can be isotopically purified with nuclei of zero spin therefore suppressing hyperfine interactions \cite{Kane_Nature98,Fodor_JPCM06}. Despite such advantages, our understanding of spin phenomena in these materials is still in its infancy. In addition, measured spin lifetimes in electrically injected Si and Ge are strongly suppressed by interfacial and degenerate doping conditions \cite{Dash_Nature09,Zhou_PRB11,Jeon_PRB11,Saito_SSC11,Jain_arxiv11,Kasahara_arxiv11,Li_nature_com11}. % giving rise to significant departure from the intrinsically attainable spin lifetimes

In this Letter we study spin properties of holes and electrons in Ge quantum wells (QWs). By measuring the circular polarization of different photoluminescence (PL) spectral peaks we quantify the spin-polarization and relaxation of optically oriented holes and electrons. The undoped and strained QWs exhibit a highly efficient spin polarization degree, yielding about 37\% and 85\% for transitions involving heavy and light holes states, respectively. The spin lifetime of electrons is found to be in the 5~ns range at 150~K and in the 0.5~ps range for holes. Our theoretical analysis shows that the spin relaxation of electrons is governed by intervalley scattering due to electron-phonon interaction. We also provide concise selection rules for the dominant phonon-assisted optical transitions. These selection rules establish a direct relation between the spin polarization of electrons and the measured circular polarization degrees of phonon-assisted spectral transitions.

Our findings are important in several regards. First, the measured spin relaxation in undoped Ge QWs is shown to be comparable to bulk Ge \cite{Guite_PRL11}, while being much longer than in ferromagnet/oxide/Ge structures in which localized states, potentially due to oxygen vacancies at the interface between the tunnel barrier and the semiconductor, mask the spin signal of the itinerant electrons \cite{Zhou_PRB11,Jeon_PRB11,Saito_SSC11,Jain_arxiv11,Kasahara_arxiv11}. Second, the extracted spin lifetime of holes is important in clarifying recent Hanle measurements in heavily-doped \textit{p}-type silicon \cite{Dash_Nature09,Gray_APL11}. The ultrafast spin relaxation of holes is governed by transitions between light and heavy holes bands. The mixed spin components of light holes lead to these enhanced rates. The magnitude of the spin-orbit coupling, as manifested by the energy difference from the split-off band, hardly plays a role in setting the scattering rate. Our findings of a ~0.5 ps hole spin lifetime are in accord with previous experimental results \cite{Hilton_PRL02,Loren_APL09}. These results may imply that localized interfacial states rather than free holes give rise to the non-negligible spin lifetime in ferromagnet/oxide/\textit{p}-type semiconductor structures. Third, in indirect band-gap semiconductors, the use of optical orientation to polarize spins (or photoluminescence to measure their polarization) has recently been analyzed theoretically \cite{Virgilio_PRB09,Pengke_PRL10,Rioux_PRB10,Cheng_PRB11}. In addition, confinement and strain in Ge QWs allows one to tune the spectral regions, resolve contributions of different hole species, and analyze indirect optical transitions.

\begin{figure}
   \centering
   \includegraphics[width=8.6cm]{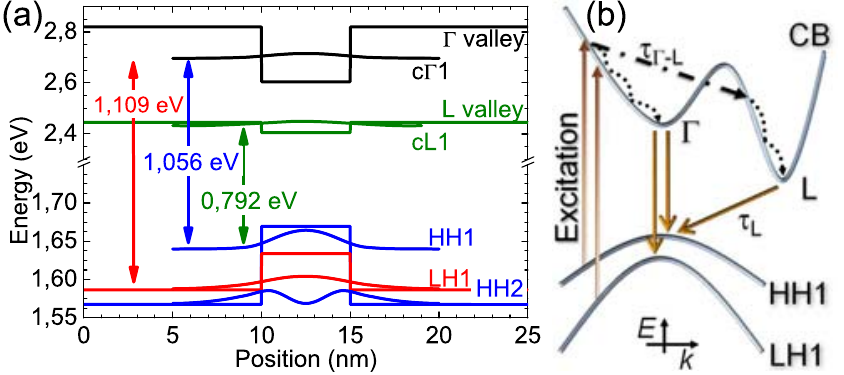}
   \caption{(color online). (a) 8-band $\mathbf{k}\cdot\mathbf{p}$ calculation of energy levels and wavefunction square amplitudes for a Ge/Si$_{0.15}$Ge$_{0.85}$ quantum well. (b) Sketch of the luminescence process. Immediately after direct gap excitation and within a timescale $\tau_{\Gamma - L}$, electrons are either scattered towards the edge of the  $\Gamma$-valley and recombine radiatively or scattered to the low energy $L$ valleys where they will recombine after $\tau_L$.} \label{fig:relax}
\end{figure}

The sample investigated in this work is a type I Ge/Si$_{0.15}$Ge$_{0.85}$ multiple quantum well (MQW) heterostructure deposited by low-energy plasma-enhanced CVD on a strain relaxed Si$_{0.1}$Ge$_{0.9}$/Si(001) graded buffer layer \cite{Rosenblad_JVST98}. The active region consists of 100 periods of a 5 nm wide Ge QW and a 10 nm thick Si$_{0.15}$Ge$_{0.85}$ barrier \cite{Bonfanti_PRB08}. An 8-band $\mathbf{k}\cdot\mathbf{p}$ modeling was performed by using NEXTNANO$^3$ \cite{nextnano}, and literature input parameters \cite{Paul_PRB08}. The resulting energy levels are shown in Fig.~1(a). This structure allows us to probe rich spin physics where spin lifetimes of holes are studied from the circular polarization degree of the  $\Gamma$~point PL via the c$\Gamma$1-HH1 and c$\Gamma$1-LH1 transitions [see Fig.~1(a)]. Similarly, spin lifetimes of electrons are studied from the $L$ point PL via the c$L$1-HH1 transition and its LA phonon replica emission. We use the sketch in Fig.~1(b) to illustrate the initial excitation of electrons in the $\Gamma$ valley, the scattering to lower energy $L$ valleys within a scattering time $\tau_{\Gamma - L}$, and the various recombination processes. %Different polarization degrees of the c$\Gamma$1-LH1 and c$\Gamma$1-HH1 lines in steady state conditions are made possible in Ge is observable in Ge thanks to a mechanism completely absent in III-V compounds, namely the scattering to the side valleys. % This phenomenon has a significant impact on the spin dynamics, since it reduces the permanence of electrons in the $\Gamma$-region to hundreds of fs, thus limiting the exciton lifetimes [23-26].

The $\Gamma$ point PL is a result of a very small portion of electrons that recombine radiatively within $\tau_{\Gamma - L}<1$~ps \cite{Loren_APL09,Lange_PRB09,Gatti_APL11}. During this ultrafast time scale, there is a net transfer of carriers from light to heavy hole bands. This transfer is enabled by the confinement and biaxial compressive strain which split the top of the valence band in the QW region. As a result of this transfer, heavy holes depolarize faster and consequently the circular polarization of the c$\Gamma$1-HH1 transition becomes smaller than that of the c$\Gamma$1-LH1 transition. After the transfer of electrons to $L$ valleys their recombination lifetime (due to radiative and non-radiative channels) is of the order of a few ns \cite{Giorgioni_JAP11}. The $L$ point luminescence will, therefore, remain circularly polarized if the spin lifetime exceeds this timescale. Two clear advantages of Ge QWs allow one to explore this rich physics by the PL. First, the recombination lifetimes in \textit{undoped} bulk IV semiconductors and SiGe heterostructures are hundreds of ns and longer \cite{Sturm_PRL91,Pankove_Book}. Unlike Ge QWs, spin polarization in these structures can be optically probed only if extremely long spin lifetimes are present. Second, the energetic proximity and ultrafast scattering of electrons from $\Gamma$ to $L$ valleys ($\Delta_{\Gamma -L}\sim 140$~meV and $\tau_{\Gamma - L} <1$~ps) allow one to resolve the spin dynamics of holes and to keep the spin polarization of electrons after scattering to $L$ valleys. Finally, other than being compatible with the mainstream Si technology, these properties also recognize Ge as an unmatched candidate in group IV photonics \cite{Michel_NaturePhotonics10,Soref_NaturePhotonics10}.

 \begin{figure}
   \centering
   \includegraphics[width=8.6cm]{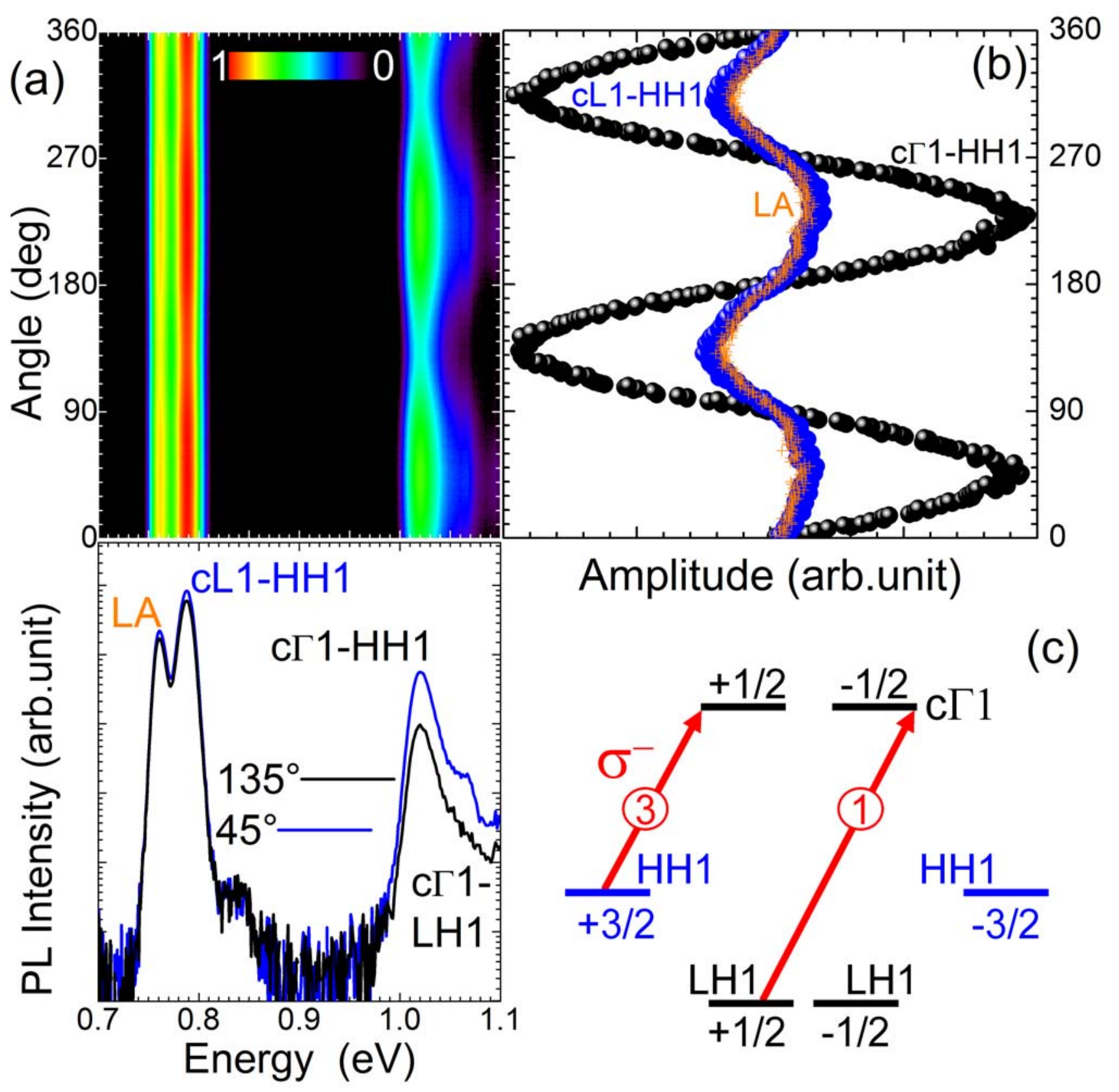}
   \caption{(color online). (a) Color-coded contour plot of the photoluminescence intensity recorded for analyzer angles ranging from 0 to 360$^{\circ}$ for the Ge/Si$_{0.15}$Ge$_{0.85}$ MQW sample at 4~K and under left-handed ($\sigma^-$) circularly polarized excitation at 1.165~eV.  The black and blue lines at the lower left panel refer to spectra resolved for analyzer angles of 135$^{\circ}$ and 45$^{\circ}$, respectively. Their intensity difference at the lower energy spectral region ($<$0.8~eV) stems from spin polarization of electrons, and at higher energies ($>$1~eV) from spin polarization of both electrons and holes. (b) Peak amplitude modulations of the c$\Gamma$1-HH1, cL1-HH1 and phonon replica emissions. (c) Optical selection rules of the $\sigma^-$ excitation. Dipole allowed transitions of heavy holes ($m_j=3/2$) are three times larger than of light holes ($m_j=1/2$). Their energy levels are split by confinement and strain. }
   \label{fig:relax}
\end{figure}

PL measurements were performed in a back-scattering geometry in a closed-cycle cryostat. A continuous wave Nd-YVO$_4$ laser, operating at 1.165~eV, was coupled to an optical retarder as circularly polarized excitation source. The laser beam was focused to a spot size having a $\sim$100~$\mu$m diameter, resulting in an excitation density in the range of 9$\times$10$^2$ - 3$\times$10$^3$~W/cm$^2$. PL polarization was then probed by the continuous rotation of a quarter-wave plate followed by a polarizer, long pass filters and a grating spectrometer equipped with an InGaAs array detector \cite{supp}. Changing the excitation from left- and right-handed circular polarization ($\sigma^-$ and $\sigma^+$), the PL signal showed the same polarization degree with opposite sign. Figure 2(a) shows a contour plot of the PL intensity at 4~K following $\sigma^-$ excitation as a function of the rotation angle of the polarization analyzer. The lower panel shows the resulting PL spectrum where emission in the $\sim$1.03~eV region is due to the c$\Gamma$1-HH1 transition. The feature at $\sim$1.06~eV is attributed to the c$\Gamma$1-LH1 excitonic recombination, superimposed onto the broad high energy tail of the main c$\Gamma$1-HH1 peak. The low-energy PL doublet, at $\sim$0.8~eV, is associated with transitions across the indirect band-gap: The high energy part is ascribed to the no-phonon emission associated with the cL1-HH1 recombination of confined carriers, while the low energy part of the doublet is the longitudinal acoustic (LA) phonon-assisted optical transition \cite{Bonfanti_PRB08}.

Figure~2(b) shows the amplitude modulation of the PL peaks for the c$\Gamma$1-HH1 (black dots), cL1-HH1 (blue dots) and the phonon replica (orange crosses) transitions. This figure reveals a sinusoidal behavior with period $\pi$, characteristic of circularly polarized emission with the same helicity of the excitation. Such observation is indeed corroborated by a full polarimetric analysis based on the measurement of the Stokes parameters $S_i$ (i=1,3) and summarized in the supplementary material \cite{supp}. Using this analysis, the circular polarization degree of the c$\Gamma$1-HH1 spectral region is $P_c = 37\% \pm 5\%$. %, i.e. higher than the maximum theoretical limit of 25\% for band-edge luminescence at the direct gap of bulk Ge, as expected from selection rules for transitions between levels with total angular momenta J=3/2 and J=1/2 \cite{Zutic_RMP04}.

The polarization analysis of the c$\Gamma$1-HH1 emission process sheds light on the hole spin relaxation process. Under $\sigma^-$ excitation at 1.165~eV, electrons are injected to $|J, J_z \rangle$ = $|1/2,1/2 \rangle$ conduction states starting from $|3/2,3/2 \rangle$ heavy hole states, as well as to $|1/2,-1/2 \rangle$ conduction states starting from $|3/2,1/2\rangle$ light hole  states [Fig 2(c)]. According to tight binding calculations \cite{Virgilio_PRB09} and data available for  III-V QWs \cite{Uenoyama_PRL90}, the overall electron polarization right after injection is expected to be $P_0$ $\approx$ 28-34\% for excitation away from the first confined HH and LH states. Under the assumption of complete HH depolarization and after thermalization towards the c$\Gamma$1 level, electrons can recombine with either $|3/2,3/2 \rangle$  or $|3/2,-3/2 \rangle$  holes emitting respectively  $\sigma^-$ or  $\sigma^+$ circularly polarized light. Assuming electron spin depolarization to be negligible, the luminescence circular polarization degree for the c$\Gamma$1-HH1 transition is $P_c \approx 0.96\cdot P_0$ \cite{Virgilio_PRB09}. The consistency of this conclusion with our experimental finding of 37\% $\pm$ 5\% validates the presence of equal numbers of heavy holes in the two sublevels, despite the fact that only $|3/2,+3/2\rangle$  holes are created upon $\sigma^-$ excitation [Fig. 2(c)]. Most importantly, the mechanisms leading to equalization of the $|3/2,-3/2 \rangle$ and $|3/2, 3/2 \rangle$ populations take place on a faster time scale than the electron scattering to $L$ valleys ($\tau_{\Gamma - L} \approx$~500~fs \cite{Lange_PRB09}).  %time despite the decrease in the valence band mixing due to strain and confinement effects. The value of $P_{circ}$ larger than the bulk limit demonstrates that the heteroepitaxial growth of Ge MQWs provides an effective means to engineer optical injection of spins.

A well-resolved c$\Gamma$1-LH1 peak with a left-handed circular polarization character is observed in Fig. 2(a). Because of its presence at the high energy tail of the c$\Gamma$1-HH1 PL band, we estimate its polarization degree by subtracting the exponential Boltzmann-like tail of the HH1 transition from the spectra obtained at the analyzer angle of $\pi$/4 and 3$\pi$/4 \cite{supp}. By doing so, we obtain a polarization as high as about 85\% $\pm$ 17\%. This result strongly suggests that LH are still polarized at the moment of electron-hole recombination. As discussed above, after $\sigma^-$ excitation only $|3/2,1/2\rangle$  light holes are created, and since the weak c$\Gamma$1-LH1 $\sigma^+$ emission is related only to $|1/2,+1/2\rangle$  electrons recombining with $|3/2,-1/2\rangle$  light holes, we conclude that the latter are not sizably present in the sample. Our PL measurements indicate that photo-generated $|3/2,+1/2 \rangle$ holes do not lose their spin orientation, but either recombine with $|1/2,-1/2\rangle$  electrons, or relax into the lower energy $|3/2,-3/2\rangle$  HH states via preserving parity scattering events \cite{Uenoyama_PRL90}. By doing so, LH can contribute to counterbalance the $|3/2,3/2\rangle$  hole population produced during the absorption, therefore depolarizing HH states. The higher energy of the LH1 band inhibits inward scattering from heavy-holes. Our measurements indicate that the spin relaxation in the LH1 band is larger than $\tau_{\Gamma - L}$.

We now focus on the indirect band-gap luminescence. The lower left panel of Fig.~2(a) shows the no-phonon (NP) cL1-HH1 peak at $\sim$0.79~eV and the LA phonon-assisted peak at $\sim$0.76~eV. In the Supplementary material we show that the polarized component of the indirect transition at 4~K has a nearly complete  $\sigma^-$ character (i.e., S3 parameter smaller than -1) \cite{supp}. The NP peak reveals a net circular polarization degree of about 8\% $\pm$ 5\%, and similarly the LA peak displays 6\% $\pm$ 5\%. We employ group theory analysis similar to Ref.~\cite{Pengke_PRL10} and find the ratio of intensities between left and right circularly polarized PL detected after propagating along the [001] crystallographic direction. Following recombination of spin-down electrons from any of the four $L$ valleys with heavy-holes, $I_{\sigma^-}:I_{\sigma_+}=3:1$ for the dominant LA assisted transition (25~meV below the band edge). For the weaker TO (36~meV) and TA (8~meV) assisted transitions the ratio is 1:3 and 1:0, respectively. The slight difference in measured polarizations of the NP ($\sim$8\%) and LA replica ($\sim$6\%) can be explained by the proximity of the weak TA transition to the NP emission region and of the weak TO transition to the LA region. Considering the aforementioned initial electron polarization of $P_0 \approx 30$\% at the $\Gamma$~valley, and the predicted LA intensity $I_{\sigma^-}^{LA}/I_{\sigma_+}^{LA}=3$ one should expect at most  $P_c \approx 15$\%. We attribute the smaller measured average value ($\sim 6\%$) to two spin-relaxation mechanisms. The first is due to electron-hole exchange interaction of confined exciton states. This mechanism plays an important role in liquid-helium temperatures and intrinsic QWs \cite{Maialle_PRB93}. The second relaxation process is attributed to the Elliott-Yafet mechanism via intervalley electron scattering by shortwave phonon modes. As explained below, this mechanism pertains to higher temperatures.

We performed a Stokes analysis of both NP and LA emissions as a function of the lattice temperature \cite{supp}. %The sign of the polarization degree has been chosen to be consistent with the sign of the Stokes parameter $S_3$, which defines the type of circular polarization.
Figure 3(a) shows that below 150~K the measured polarization of the indirect band-gap luminescence in Ge/SiGe MQWs is nearly independent of temperature \cite{footnote}. This observation is a clear indication that the optically oriented spins withstands the ultrafast relaxation to the $L$ valleys as well as the dwell time of electrons in the $L$ valleys before recombination. We make use of recent PL decay measurements of the recombination lifetime of similar samples  \cite{Giorgioni_JAP11}. By taking an average of Ge QW widths of 3.8~nm and 7.3~nm, the NP decay times are only slightly decreasing with temperature while kept below 10~ns between 14 and 300~K. The data is shown by the flatter (blue) curve in Fig. 3(b).

To complete the picture we have also performed spin relaxation analysis due to electron-phonon scattering. The sharply decreasing (black) curve in Fig.~3(b) shows the resulting spin relaxation time ($\tau_s$) in intrinsic Ge. Simulations were made using a spin-dependent $\mathbf{k}\cdot\mathbf{p}$ expansion at the vicinity of the $L$ point. Similar to the spin-dependent expansion near the $X$ point of silicon \cite{Pengke_PRL11}, such modeling provides useful insights. The detailed theory will be shown in a different paper and here we mention key findings.  Around 150~K $\tau_s \approx 5$~ns is comparable with the recombination lifetime, $\tau_L$, in very good agreement with the reduction in circular polarization of the NP emission beyond this temperature range, $P_{c} = P_{c}^0 / (1+ \tau_L/\tau_s$) \cite{Parsons_PRL69}. The spin relaxation is governed by \textit{intervalley} electron-phonon scattering in a wide temperature range (T$>$30~K) \cite{Tang_Arxiv11}.  This dominating mechanism involves shortwave phonon modes whose wavevector connect centers of different $L$ valleys (i.e., phonon modes near the $X$ point). Dominant contributions result from the $X_4$ and $X_1$ symmetries where their phonon energies are $\approx$30~meV. The temperature dependence of this spin relaxation mechanism is governed by the Bose-Einstein distribution of these phonon modes.   Using the symmetries of the $L$ point space group ($D_{3d}$), the intervalley spin relaxation rate can be shown to depend on the spin-orbit coupling \textit{between conduction bands and not between conduction and valence bands}. At very low temperatures the population of $X$ point phonon modes is negligible and as a result the \textit{intravalley} spin relaxation rate can exceed the \textit{intervalley} rate.  The former, originally predicted by Yafet to have a $T^{7/2}$ temperature dependence \cite{Yafet_SSP63}, results in spin lifetimes of the order of a few $\mu$s or longer at low temperatures. Practically, however, such a long timescale will be easily masked by localization (freeze-out conditions), electron-hole exchange, or doping effects.

 \begin{figure}
   \centering
   \includegraphics[width=8.6cm,height=5cm]{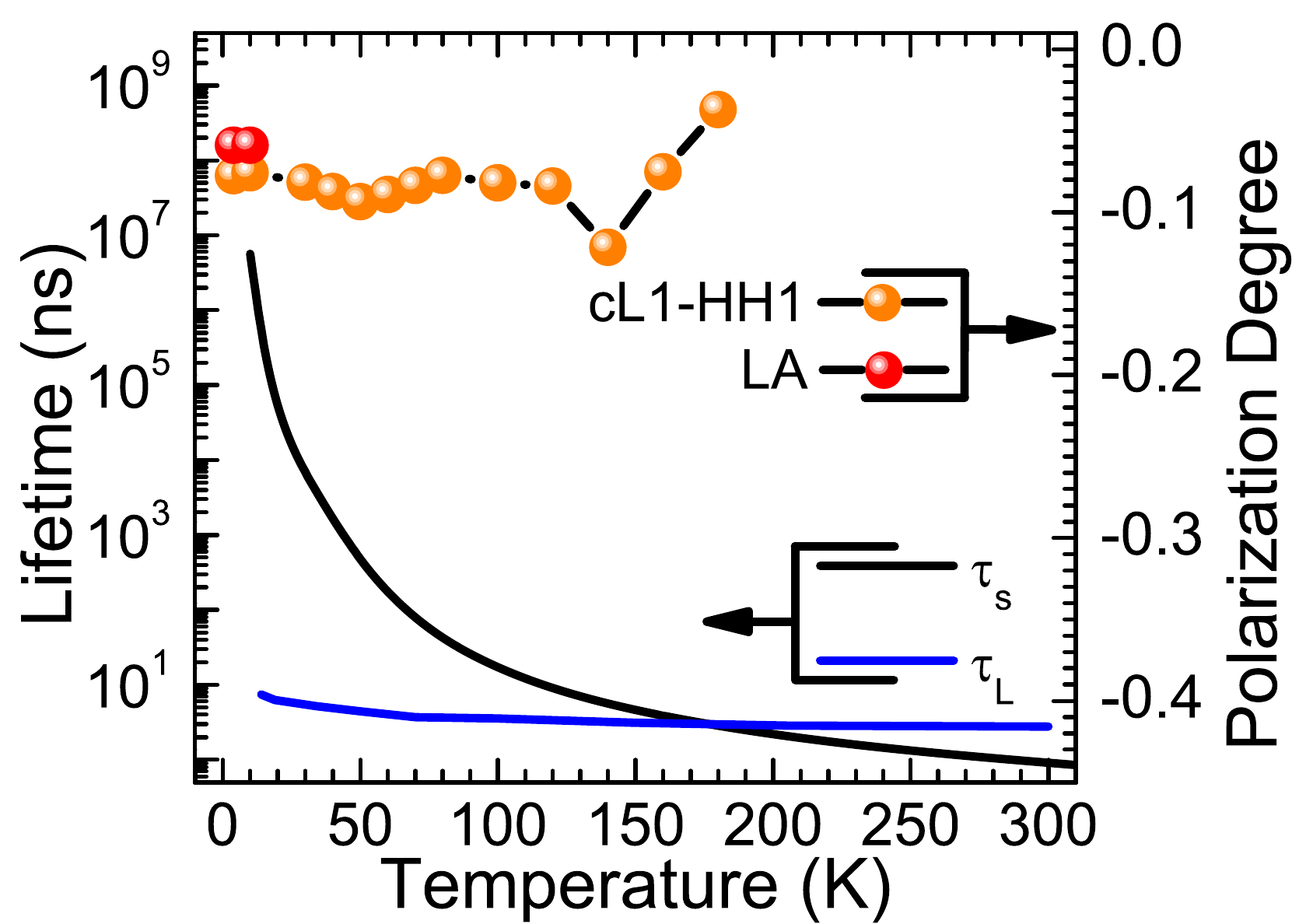}
   \caption{(color online). \textit{dots}: Measured temperature dependence of the circular polarization degree at the peak region of the no-phonon (cL1-HH1) emission (orange dots). Results of the LA phonon-assisted emission (red dots) are shown at T$<$20~K where it is well resolved from the cL1-HH1 peak. \textit{black line}: Modeled temperature dependence of the electron spin lifetime due to electron-phonon interaction. \textit{blue line}: Measured recombination lifetime due to radiative and non-radiative channels in a $\sim$5~nm wide Ge/SiGe QW (after \cite{Giorgioni_JAP11}).} \label{fig:relax}
\end{figure}

In conclusion, we have demonstrated efficient and robust optical spin orientation in Ge/SiGe MQWs. The photoluminescence from Ge quantum wells is shown to be an efficient tool in studying the spin lifetimes of holes and electrons. The high polarization degrees of the spectral region suggest that, after optical orientation carriers do not completely lose their spin memory, when they cool down or experience intervalley scattering. These results are particularly exciting for spintronic applications, since an optical spin injection scheme based upon the  $\Gamma$-to-L spin transfer bypasses detrimental interfacial effects of all-electrical ferromagnet/oxide/Ge injection schemes. In addition, efficient optical emission and spin polarization at the direct gap of Ge-based heterostructures open up a unique possibility to merge the potential of both photonics and spintronics on the well-established silicon platform. With this respect, the use of Ge heterostructures allows one to tune the emitted photon wavelength by controlling the well width and strain levels.

This work was supported by the PRIN Project No. 20085JEW12, by Regione Lombardia through Dote ricercatori, by AFOSR Contract No. FA9550-09-1-0493, and by NSF Contract No. ECCS-0824075. The authors would like to thank E. Gatti and D. Chrastina for their support in  sample design and preparation and A. Rastelli for the use of xrsp software.


\begin{thebibliography}{99}
\bibitem{footnote} By increasing the temperature the band gap shrinks and above 150~K the indirect gap emission starts to leak out of the spectral range   of   our apparatus.
\end{thebibliography}
\end{document}